\documentclass[sigconf]{acmart}

\AtBeginDocument{%
  }

\setcopyright{acmlicensed}
\copyrightyear{2018}
\acmYear{2018}
\acmDOI{XXXXXXX.XXXXXXX}
\acmConference[Conference acronym 'XX]{Make sure to enter the correct
  conference title from your rights confirmation email}{June 03--05,
  2018}{Woodstock, NY}
\acmISBN{978-1-4503-XXXX-X/2018/06}

\startPage{1}

\usepackage{multirow}
\usepackage{xcolor}
\usepackage{booktabs}
\usepackage{graphicx}
\usepackage{enumitem}
\usepackage{algorithmic}
\usepackage{algorithm}
\usepackage{subcaption}
\begin{document}

\title{PIT: A Dynamic Personalized Item Tokenizer for End-to-End Generative Recommendation}

\author{Huanjie Wang}
\orcid{0009-0009-6046-2008}
\affiliation{%
    \institution{Beijing University of Posts and Telecommunications}
    \city{Beijing}
    \country{China}
}
\email{wanghuanjie@bupt.edu.cn}

\author{Xinchen Luo}
\affiliation{%
    \institution{Kuaishou Technology}
    \city{Beijing}
    \country{China}
}
\email{luoxinchen@kuaishou.com}

\author{Honghui Bao}
\affiliation{%
    \institution{Kuaishou Technology}
    \city{Beijing}
    \country{China}
}
\email{honghuibao2000@gmail.com}

\author{Zixing Zhang}
\affiliation{%
    \institution{Kuaishou Technology}
    \city{Beijing}
    \country{China}
}
\email{zhangzixing@kuaishou.com}

\author{Lejian Ren}
\affiliation{%
    \institution{Kuaishou Technology}
    \city{Beijing}
    \country{China}
}
\email{renlejian@kuaishou.com}

\author{Yunfan Wu}
\affiliation{%
  \institution{Institute of Computing Technology, Chinese Academy of Sciences}
  \city{Beijing}
  \country{China}
}
\email{wuyunfan15@mails.ucas.ac.cn}

\author{Hongwei Zhang}
\affiliation{%
    \institution{Kuaishou Technology}
    \city{Beijing}
    \country{China}
}
\email{zhanghongwei08@kuaishou.com}

\author{Liwei Guan}
\affiliation{%
    \institution{Kuaishou Technology}
    \city{Beijing}
    \country{China}
}
\email{guanliwei@kuaishou.com}

\author{Guang Chen}
\authornote{Corresponding author.}
\affiliation{%
    \institution{Beijing University of Posts and Telecommunications}
    \city{Beijing}
    \country{China}
}
\email{chenguang@bupt.edu.cn}

\renewcommand{\shortauthors}{Wang et al.}


\begin{abstract}
Generative Recommendation has revolutionized recommender systems by reformulating retrieval as a sequence generation task over discrete item identifiers. Despite the progress, existing approaches typically rely on static, decoupled tokenization that ignores collaborative signals. While recent methods attempt to integrate collaborative signals into item identifiers either during index construction or through end-to-end modeling, they encounter significant challenges in real-world production environments. Specifically, the volatility of collaborative signals leads to unstable tokenization, and current end-to-end strategies often devolve into suboptimal two-stage training rather than achieving true co-evolution. To bridge this gap, we propose \textbf{PIT}, a dynamic \underline{P}ersonalized \underline{I}tem \underline{T}okenizer framework for end-to-end generative recommendation, which employs a co-generative architecture that harmonizes collaborative patterns through collaborative signal alignment and synchronizes item tokenizer with generative recommender via a co-evolution learning. This enables the dynamic, joint, end-to-end evolution of both index construction and recommendation. Furthermore, a one-to-many beam index ensures scalability and robustness, facilitating seamless integration into large-scale industrial deployments. Extensive experiments on real-world datasets demonstrate that PIT consistently outperforms competitive baselines. In a large-scale deployment at Kuaishou, an online A/B test yielded a substantial \textbf{0.402\%} uplift in App Stay Time, validating the framework's effectiveness in dynamic industrial environments.
\end{abstract}

\ccsdesc[500]{Information systems~Recommender systems}

\keywords{Short-Video Recommendation, Generative Recommendation, LLMs for Recommendation, Item Tokenization, Learnable Tokenizer}

\maketitle


\section{Introduction}
\label{sec:introduction}

As the cornerstone of modern digital platforms, recommender systems must efficiently process billions of interactions to deliver personalized content within an ever-expanding item universe. Traditionally, the dominant paradigm follows a ``Retrieve-then-Rank'' funnel~\cite{huang2013LearningDeepStructured,covington2016DeepNeuralNetworks,cheng2016WideDeepLearning, guo2017DeepFMFactorizationMachineBased,zhou2018DeepInterestNetwork}, which is inherently limited by the accuracy-efficiency trade-off in approximate search of vector retrieval and the heavy memory footprint required for storing dense indices~\cite{johnson2017BillionscaleSimilaritySearch, malkov2018EfficientRobustApproximate}. Recently, Generative Recommendation (GR) has emerged as a disruptive paradigm to overcome these limitations~\cite{wang2024GenerativeRecommendationNextgeneration, li2024LargeLanguageModelsa, li2025SurveyGenerativeRecommendationa}. Unlike cascaded pipelines that suffer from optimization inconsistencies and fragmented computation, GR enables holistic optimization within a unified architecture, significantly reducing communication overhead~\cite{zhou2025OneRecTechnicalReport}. Furthermore, by replacing atomic identifiers with semantic identifiers, it facilitates knowledge transfer across items and demonstrates superior parameter efficiency at scale\cite{rajput2023RecommenderSystemsGenerative, zhou2025OneRecTechnicalReport}. Adopting generative retrieval as its core mechanism, GR reformulates recommendation as a sequence generation task over discrete item identifiers~\cite{tay2022TransformerMemoryDifferentiable,wang2023NeuralCorpusIndexer}. This shift unifies indexing and recommendation into a single differentiable Transformer~\cite{vaswani2023AttentionAllYou} model, demonstrating promising potential in both efficiency and deep semantic interaction~\cite{rajput2023RecommenderSystemsGenerative, zhou2025OneRecTechnicalReport, liu2025OneRecThinkInTextReasoning}.

However, a critical bottleneck remains in the tokenization process, the mechanism of assigning discrete identifiers to items in GR. Existing approaches~\cite{rajput2023RecommenderSystemsGenerative, zheng2024AdaptingLargeLanguage, wang2024EAGERTwoStreamGenerative, si2024GenerativeRetrievalSemantic, liu2025DiscRecDisentangledSemanticCollaborativea} typically adopt a disjoint, two-stage pipeline: (1) static tokenization, where quantizers (e.g., RQ-VAE~\cite{lee2022AutoregressiveImageGeneration}) map items to static identifiers using hierarchical clustering~\cite{wang2024ContentBasedCollaborativeGeneration, wang2024EAGERTwoStreamGenerative, si2024GenerativeRetrievalSemantic} or content reconstruction~\cite{rajput2023RecommenderSystemsGenerative, wang2025LearnableItemTokenization, liu2025GenerativeRecommenderEndtoEnd, lin2025UnifiedSemanticID}; and (2) generative recommender learning~\cite{tay2022TransformerMemoryDifferentiable,wang2023NeuralCorpusIndexer, rajput2023RecommenderSystemsGenerative, zhou2025OneRecTechnicalReport}, where the generative model is trained to predict these fixed identifiers as immutable labels. This separation results in a fundamental ``semantic gap'': the representations optimized for reconstructing item content inherently deviate from those that are the most predictable by a generative recommender based on historical interactions \cite{wang2025LearnableItemTokenization}.

To surmount the limitations of these disjoint pipelines, the research focus has shifted towards learnable~\cite{wang2025LearnableItemTokenization} and end-to-end~\cite{liu2025GenerativeRecommenderEndtoEnd, bai2025BiLevelOptimizationGenerative} tokenization strategies. The core motivation is to eliminate the barrier between index construction and recommendation, enabling a unified optimization landscape where the item tokenizer is explicitly guided by recommendation rather than mere reconstruction fidelity~\cite{liu2025GenerativeRecommenderEndtoEnd,bai2025BiLevelOptimizationGenerative}. Theoretically, this seamless integration promises to refine item representations through deep fusion with collaborative signals, achieving mutual enhancement between the tokenizer and the generative recommender. However, adapting these academic advances into real-world production environments remains a formidable challenge~\cite{zou2025SurveyRealWorldRecommender}, primarily due to the severe distribution shifts in collaborative signals driven by real-time interactions~\cite{zhang2023InvariantCollaborativeFiltering, yang2023GenericLearningFramework, zhang2024RobustCollaborativeFiltering, zou2025SurveyRealWorldRecommender}. In such highly dynamic industrial environments, interaction patterns and their corresponding optimal collaborative representations undergo continuous evolution~\cite{yang2023GenericLearningFramework, zhang2023InvariantCollaborativeFiltering, zou2025SurveyRealWorldRecommender}. This volatility is incompatible with the optimization mechanisms of existing end-to-end methods~\cite{liu2025GenerativeRecommenderEndtoEnd, bai2025BiLevelOptimizationGenerative}, which typically rely on rigid alternating optimization or a phased training schedule where the tokenizer is frozen in the later stages to mitigate instability. Consequently, these approaches fail to adapt to rapid concept drifts, making it impractical to maintain a stable and synchronized index within a large-scale online learning framework, thereby obstructing their widespread industrial deployment.

To address these challenges, we argue that index construction and user preference learning should be a true co-evolutionary process. In this paper, we propose \textbf{PIT}, a dynamic \underline{P}ersonalized \underline{I}tem \underline{T}okenizer framework for end-to-end generative recommendation. Specifically, PIT comprises three components: First, we propose a \textit{co-generative architecture} comprising an \textit{Item-to-Token} model for indexing, a \textit{User-to-Token} model for generative recommendation, and a \textit{collaborative signal alignment} component. The latter is a DIN-based~\cite{zhou2018DeepInterestNetwork} auxiliary task designed to reconcile the discrepancy between semantic and collaborative signals. Second, we propose a \textit{co-evolution learning} paradigm centered on a user-guided \textit{minimum-loss selection} mechanism. This mechanism enables the concurrent optimization of both Item-to-Token and User-to-Token processes by dynamically selecting the identifier that minimizes prediction loss, thereby ensuring their seamless synchronization. Finally, a one-to-many \textit{beam index} is employed to maintain multiple valid identifiers per item, which not only mitigates the fragility of single-path indexing against distribution shifts but also supports seamless and dynamic index updates during industrial streaming training

In summary, our main contributions are as follows:
\begin{itemize}[nosep,leftmargin=*]
\item We propose PIT, a dynamic end-to-end generative framework that unifies item tokenization and recommendation into a co-generative architecture, ensuring identifiers are optimized specifically for collaborative signals.
\item We introduce co-evolution learning via a minimum-loss selection mechanism and one-to-many beam index, enabling concurrent optimization of identifiers and models to mitigate instability in industrial streams.
\item We validate PIT through both extensive offline experiments and a large-scale online A/B test, where it achieved a 0.402\% uplift in App Stay Time, demonstrating its robustness and business value in a hundred-million-scale environment.
\end{itemize}

\section{Related Work} 
\label{sec:related work}
\noindent \textbf{Generative Recommendation.} 
The shift towards Generative Recommendation has transitioned the recommendation paradigm from traditional vector similarity searches~\cite{johnson2017BillionscaleSimilaritySearch, malkov2018EfficientRobustApproximate} to an autoregressive sequence-to-sequence task~\cite{sutskever2014SequenceSequenceLearning, wang2024GenerativeRecommendationNextgeneration, li2024LargeLanguageModelsa, li2025SurveyGenerativeRecommendationa}. Central to this evolution is the design of item identifiers, which serve as the bridge between raw item features and the model's discrete output space. Early works such as P5~\cite{geng2023RecommendationLanguageProcessing} utilize text-based tokens to predict the next item. Building upon this, TIGER~\cite{rajput2023RecommenderSystemsGenerative} introduces the concept of semantic identifiers (SIDs), which represent items through discrete identifiers quantized from content embeddings, enabling the model to better capture item-specific semantics. To accommodate extensive item corpora, recent industrial advancements such as OneRec~\cite{zhou2025OneRecTechnicalReport} demonstrate the efficacy and scalability of GR by deploying unified SIDs within large-scale commercial infrastructures. Furthermore, LC-Rec~\cite{zheng2024AdaptingLargeLanguage} and OneRec-Think~\cite{liu2025OneRecThinkInTextReasoning,zhou2025OpenOneRecTechnicalReport} explicitly integrate SIDs into the Large Language Model (LLM)~\cite{vaswani2023AttentionAllYou, radfordImprovingLanguageUnderstanding} modality, treating items as a distinct native vocabulary. These methods~\cite{wang2024GenerativeRecommendationNextgeneration, li2024LargeLanguageModelsa, li2025SurveyGenerativeRecommendationa} demonstrate that autoregressive generation can achieve deeper semantic matching than conventional recommendation paradigms such as dual-encoder~\cite{huang2013LearningDeepStructured,covington2016DeepNeuralNetworks}. However, the performance of these recommenders remains critically dependent on the quality of the SIDs used as generation targets.

\vspace{0.5em}
\noindent \textbf{Item Tokenization.} 
The core of Generative Recommendation lies in item tokenization, which maps items into a discrete identifier space. Existing strategies for constructing these tokenizers can be broadly categorized into static and learnable approaches~\cite{wang2024GenerativeRecommendationNextgeneration, li2024LargeLanguageModelsa, li2025SurveyGenerativeRecommendationa}.
(1) \textbf{Static Tokenization.} Methods like TIGER~\cite{rajput2023RecommenderSystemsGenerative} and LC-Rec~\cite{zheng2024AdaptingLargeLanguage} adopt a decoupled two-stage pipeline: items are first tokenized offline based on content features (e.g., RQ-VAE~\cite{lee2022AutoregressiveImageGeneration, rajput2023RecommenderSystemsGenerative}), which freezes the identifier space prior to generative recommender training. However, such a decoupled design suffers from a misalignment between content and collaborative signals~\cite{wang2025LearnableItemTokenization}. Since tokenizers optimize for reconstruction rather than collaborative signals, behaviorally distinct items might be assigned identical identifiers due to content similarity, creating a ``semantic gap'' that limits precision~\cite{wang2025LearnableItemTokenization}.
(2) \textbf{Learnable Tokenization.} To bridge this gap, research has shifted towards learnable and end-to-end tokenization to inject collaborative signals into the index. LETTER~\cite{wang2025LearnableItemTokenization} employs collaborative regularization to optimize the codebook during tokenizer training. Moving towards more complex optimization, ETEGRec~\cite{liu2025GenerativeRecommenderEndtoEnd} achieves end-to-end tokenization but relies on an alternating optimization strategy, iteratively updating the tokenizer and generative recommender. This strategy increases training complexity, and the requirement of freezing the tokenizer in later stages limits its adaptability to real-time streaming data. 
(3) \textbf{Our Approach.} Reviewing the existing works on generative recommendation, we observe that current paradigms often fail to achieve true one-stage end-to-end training. Furthermore, their limited adaptability to large-scale industrial systems remains a critical bottleneck. To address these deficiencies, we propose a novel framework that enables the dynamic co-evolution of the tokenizer and the generative recommender, facilitating a seamless one-stage industrial pipeline, yielding superior recommendation performance.

\section{Methodology}
\label{sec:methodology}
In this section, we first formulate the generative recommendation problem and the definitions of tokenization in Section~\ref{sec:formulation}. Subsequently, we detail the \textit{co-generative
architecture} comprising the \textit{collaborative signal alignment}, \textit{Item-to-Token} and \textit{User-to-Token} models in Section~\ref{sec:architecture}. In Section~\ref{sec:learning} we introduce a \textit{co-evolution learning strategy}, which utilizes \textit{minimum-loss selection} mechanism to align index construction with user preferences. Then, in Section~\ref{sec:dynamic index}, we elaborate on the dynamic \textit{beam index}, which enables an inverse mapping from multiple SIDs to a single item. Finally, in Section~\ref{sec:system deployment}, we describe the practical online deployment and operational workflow of PIT within a large-scale industrial infrastructure.

\subsection{Problem Formulation}
\label{sec:formulation}
Let $\mathcal{U}$ and $\mathcal{I}$ denote the sets of users and items, respectively. For each user $u \in \mathcal{U}$, we represent their historical interaction history as a sequence $H_u = \{i_1, i_2, \dots, i_n\}$, where $i_t \in \mathcal{I}$. Crucially, each item $i \in \mathcal{I}$ is intrinsically characterized by a comprehensive feature set $X_i$, which integrates its unique item identifier with heterogeneous multimodal signals, such as textual descriptions, cover images, and other categorical attributes.

\vspace{0.5em}
\textbf{Generative Recommendation}. Following the standard GR paradigm~\cite{rajput2023RecommenderSystemsGenerative}, we represent each item $i$ as a sequence of discrete tokens $c_i = (c_{i,1}, c_{i,2}, \dots, c_{i,L})$ from a codebook $\mathcal{V}$ of size $K$. The goal of the generative recommender is to learn a probabilistic model $P_\theta(c_i | H_u)$ that autoregressively generates the SID of the target item given the user context:
\begin{equation}
    P_\theta(c_i | H_u) = \prod_{l=1}^{L} P_\theta(c_{i,l} | c_{i, <l}, H_u),
\end{equation}
where $c_{i, <l}$ denotes the prefix of the token sequence generated.

\vspace{0.5em}
\textbf{Item Tokenization}. In conventional approaches, $c_i$ is derived via a static function $f_{static}: X_i \to c_i$ fixed before generative recommender training. In our framework, we treat the assignment of $c_i$ as a dynamic process governed by a learnable tokenizer parameterized by $\phi$.

\subsection{Co-Generative
Architecture}
\label{sec:architecture}
\begin{figure*}[ht]
    \centering
    \includegraphics[width=\textwidth]{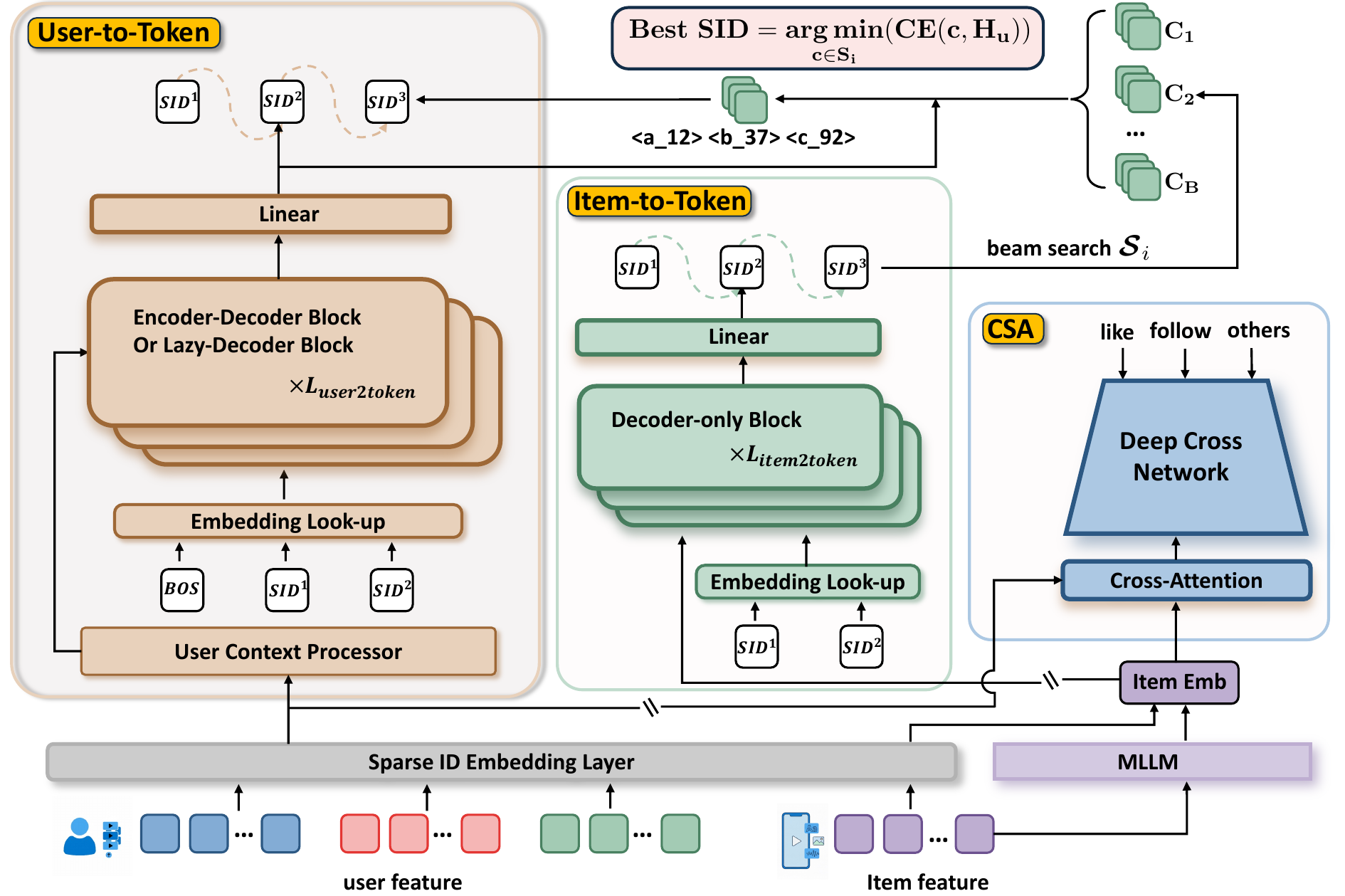}
    \caption{Illustration of the PIT framework comprising co-generative architecture when using   user-guided minimum-loss selection mechanism in co-evolution learning.}
    \label{fig:framework_overview}
\end{figure*}

Our framework follows a co-generative architecture, as shown in Figure~\ref{fig:framework_overview}, which consists of three primary components: (1) collaborative signal alignment, which enriches item representations by capturing latent user-item interactions; (2) An Item-to-Token tokenizer that yields dynamic semantic identifiers; and (3) the User-to-Token recommender for generative recommendation.

\subsubsection{collaborative signal alignment (CSA)}
We introduce an auxiliary multi-behavior prediction task to inject multi-faceted collaborative signals directly into the item embedding $X_i$. 
Furthermore, $X_i$ is enriched with multimodal embedding representations (e.g., visual and textual features), enabling the model to capture both intrinsic semantic attributes and collaborative user behaviors simultaneously. We employ a Deep Interest Network~\cite{zhou2018DeepInterestNetwork} structure as the auxiliary module. It takes the user behavior sequence $H_u$ and the target item representation $X_i$ as input. An attention mechanism computes the relevance between the target item and historical behaviors to predict the interaction probability $\hat{y}^{b}_{u,i}$ for behavior $b$:
\begin{equation}
    \hat{y}^{b}_{u,i} = \text{DIN}(\text{sg}(H_u), X_i),
\end{equation}
where $\text{sg}(\cdot)$ denotes the stop-gradient operator; $b \in \mathcal{B}$ represents a specific user behavior (e.g., clicking or liking) from the multi-faceted behavior set $\mathcal{B}$. The term $y^b_{u,i} \in \{0, 1\}$ is the ground-truth label indicating whether user $u$ performed behavior $b$ on item $i$. Crucially, we detach $H_u$ to prevent the auxiliary binary classification objective from interfering with the representation space of the main generative recommender model, ensuring the user context embeddings remain focused on the generation task. 

\subsubsection{Item-to-Token Model (Item Tokenizer)}
The Item-to-Token model, parameterized by $\phi$, acts as the index generator. Given the relative simplicity of the Item-to-Token task, we employ a lightweight decoder-only Transformer to maintain efficiency during iterative co-evolution. Functionally, the model conditions the generation on the collaborative and multimodal item representation learned in collaborative signal alignment. We use the detached item embedding $\text{sg}(X_i)$ as the initial BOS token for the Transformer, which then recursively predicts the discrete SID sequence. The probability is formulated as:
\begin{equation}
    P_\phi(c_i | \text{sg}(X_i)) = \prod_{l=1}^{L} P_\phi(c_{i,l} | c_{i, <l}, \text{sg}(X_i)).
\end{equation}
We explicitly detach $X_i$ here because the goal of the Item-to-Token is to generate tokens that describe the existing collaborative and multimodal embedding, not to modify the embedding to make token generation easier. This forces the generated SIDs to faithfully reflect the collaborative signals injected by the CSA.

\subsubsection{User-to-Token Model (Generative Recommender)}
The User-to-Token model, parameterized by $\theta$, learns to capture user preferences. It takes the interaction history $H_u$ as input and predicts the target item's SID sequence. We adopt either an encoder-decoder~\cite{cho2014LearningPhraseRepresentations, sutskever2014SequenceSequenceLearning} or a lazy-decoder~\cite{zhou2025OneRecV2TechnicalReport} as the base architecture for the User-to-Token:
\begin{equation}
    P_\theta(c | H_u) = \prod_{l=1}^{L} P_\theta(c_l | c_{<l}, H_u).
\end{equation}
The User-to-Token model primarily serves as the generative recommender. Beyond its inference capability, it acts as a critical evaluator during the co-evolution learning. By calculating the cross-entropy loss for various candidate SID sequences, the model identifies the optimal SID that is most congruent with the specific user's preferences. This selection process determines the most effective optimization direction.

\subsection{Co-Evolution Learning}\label{sec:learning}
To achieve a truly end-to-end update of both the tokenizer and generative recommender based on the co-generative architecture, we adopt a curriculum learning~\cite{bengio2009CurriculumLearning, soviany2022CurriculumLearningSurvey} strategy. This approach begins with a warm-up phase to learn from existing SID assignments, establishing robust foundational representations. Subsequently, the model transitions seamlessly into a user-guided dynamic evolution phase driven by a minimum-loss selection mechanism. Throughout this process, the collaborative signal alignment auxiliary task continuously refines item embeddings.
\subsubsection{Training Objectives}
\label{sec:objectives}
We define three atomic loss functions to modularly optimize the collaborative alignment, Item-to-Token, and User-to-Token components:
\begin{itemize}[leftmargin=*, itemsep=4pt]
    \item \textbf{Collaborative Loss ($\mathcal{L}_{xtr}$)}: Refines item embeddings $X_i$ via a multi-task Binary Cross-Entropy (BCE) loss on multi-faceted behaviors $b \in \mathcal{B}$:
    \begin{equation}
        \mathcal{L}_{xtr} = - \sum_{b \in \mathcal{B}} \sum_{(u, i) \in \mathcal{D}^b} \left( y_{u, i}^b \log \hat{y}_{u, i}^b + (1 - y_{u, i}^b) \log (1 - \hat{y}_{u, i}^b) \right).
    \end{equation}
    \item \textbf{Tokenizer Loss ($\mathcal{L}_{item}$)}: Optimizes the Item-to-Token tokenizer ($\phi$) to generate the target SID sequence $c$ conditioned on the collaborative item representation:
    \begin{equation}
        \mathcal{L}_{item}(c) = - \sum_{l=1}^{L} \log P_\phi(c_l \mid c_{<l}, \text{sg}(X_i)).
    \end{equation}
    \item \textbf{Recommender Loss ($\mathcal{L}_{user}$)}: Updates the User-to-Token recommender ($\theta$) to predict the SID sequence based on user history $H_u$:
    \begin{equation}
        \mathcal{L}_{user}(c) = - \sum_{l=1}^{L} \log P_\theta(c_l \mid c_{<l}, H_u).
    \end{equation}
\end{itemize}
These objectives are combined during the warm-up and dynamic evolution phases to achieve end-to-end co-evolution learning.

\subsubsection{Phase 1: Warm-up with Collaborative Signals}
Directly learning discrete SIDs from random initialization often leads to index instability. To establish foundational recognition capabilities, we facilitate Item-to-Token and User-to-Token learning by leveraging an existing tokenizer (e.g., RQ-VAE or RQ-Kmeans). By training all three modules simultaneously, the joint objective is:
\begin{equation}
    \mathcal{L}_{warm} = \mathcal{L}_{user}(\bar{c}_i) + \lambda_1 \mathcal{L}_{item}(\bar{c}_i) + \lambda_2 \mathcal{L}_{xtr},
\end{equation}
where $\bar{c}_i$ denotes the pre-defined SID sequence of item $i$ generated by the existing tokenizer, and $\mathcal{L}_{user}$ and $\mathcal{L}_{item}$ represent the negative log-likelihood of generating $\bar{c}_i$ for the User-to-Token and Item-to-Token, respectively. This phase ensures the models possess basic representational capabilities, with $X_i$ pre-encoding collaborative patterns prior to dynamic evolution.

\subsubsection{Phase 2: User-Guided Dynamic Evolution}
After the warm-up phase, our framework employs a dynamic generation-selection loop, as shown in Figure~\ref{fig:framework_overview}, which iteratively executes the following three steps for each mini-batch (see Appendix~\ref{app:online_training} for details):

\textbf{Step 1: Candidate Generation via Beam Search.}
For a target item $i$, the Item-to-Token proposes potential SIDs based on the item embedding $X_i$. Using beam search, we generate a set of $B$ diverse candidate SID sequences $\mathcal{S}_i$:
\begin{equation}
    \mathcal{S}_i = \text{BeamSearch}(P_\phi(\cdot | \text{sg}(X_i)), B).
\end{equation}

\textbf{Step 2: Minimum-Loss Selection.}
The candidate set $\mathcal{S}_i$ is passed to the User-to-Token model, which functions as a critic to measure the compatibility of each candidate with the user history $H_u$. Specifically, we identify the optimal identifier $c^*_i$ by minimizing the cross-entropy loss in an autoregressive manner:
\begin{equation}
    c^*_i = \underset{c \in \mathcal{S}_i}{\arg\min} \left( - \frac{1}{L} \sum_{l=1}^L \log P_\theta(c_l | c_{<l}, H_u) \right).
\end{equation}

\textbf{Step 3: Joint Optimization.}
The selected sequence $c^*_i$ serves as the dynamic pseudo-label for the current iteration. We update the entire framework via a unified objective:
\begin{equation}
    \mathcal{L}_{dynamic} = \mathcal{L}_{user}(c^*_i) + \alpha \mathcal{L}_{item}(c^*_i) + \beta \mathcal{L}_{xtr}
\end{equation}
Specifically, the gradients flow to jointly optimize the framework: $\mathcal{L}_{user}$ updates the User-to-Token parameters $\theta$ to align user representations with the best-selected token $c^*_i$ continuously refining the user representation to adapt to emerging interests in the data stream; $\mathcal{L}_{item}$ refines the Item-to-Token parameters $\phi$ to generate SIDs that fuse collaborative signals and multimodal representations, conditioned on $\mathcal{L}_{user}(c^*_i)$ is sufficiently small to ensure the selected $c^*_i$ carries substantial semantic relevance and represents a valid optimization direction aligned with true user preferences; and $\mathcal{L}_{xtr}$ refines the item embeddings $X_i$ along with DIN parameters to adapt to evolving collaborative patterns, thereby providing enhanced signals for the next Item-to-Token stage.
\begin{figure}[t]
    \centering
    \includegraphics[width=\linewidth]{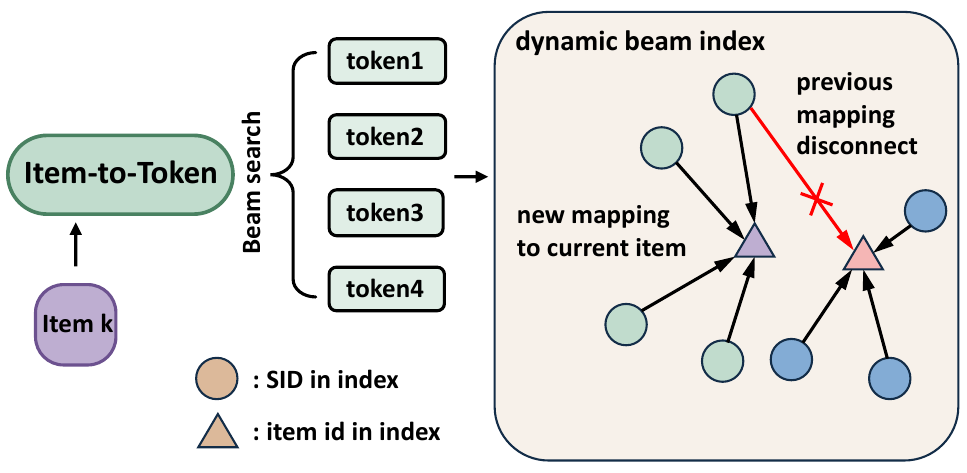}
    \caption{The mechanism of beam index which shows how PIT dynamically maps multiple SIDs and item identifiers}
    \label{fig:dynamic muti token index}
\end{figure}
\subsection{Dynamic Beam Index}
\label{sec:dynamic index}
Unlike existing generative recommendation methods that rely on static, one-to-one indexing, we propose the \textit{beam index}, a scheme that reformulates the mapping between items and SIDs as a dynamic, one-to-many relationship.
Within this scheme, item-to-SID associations are continuously optimized. For each item, a candidate set $\mathcal{S}_i = \{s_1, s_2, \dots, s_B\}$ is generated via item-to-token beam search. We refine these $B$ indices using a loss-based competitive weighting strategy: if a candidate overlaps with an existing index, its weight is updated via a momentum-based moving average of the losses; otherwise, the current loss is directly assigned as the weight. By selecting the top-$B$ weighted indices, beam index reconfigures the index topology, pruning sub-optimal identifiers to enhance collaborative signals integration. Our framework leverages the latest synchronized indices from the training phase to execute downstream tasks.(see Appendix~\ref{app:beam_index} for details).

\subsection{System Deployment}
\label{sec:system deployment}
\begin{figure}[b]
    \centering
    \includegraphics[width=\linewidth]{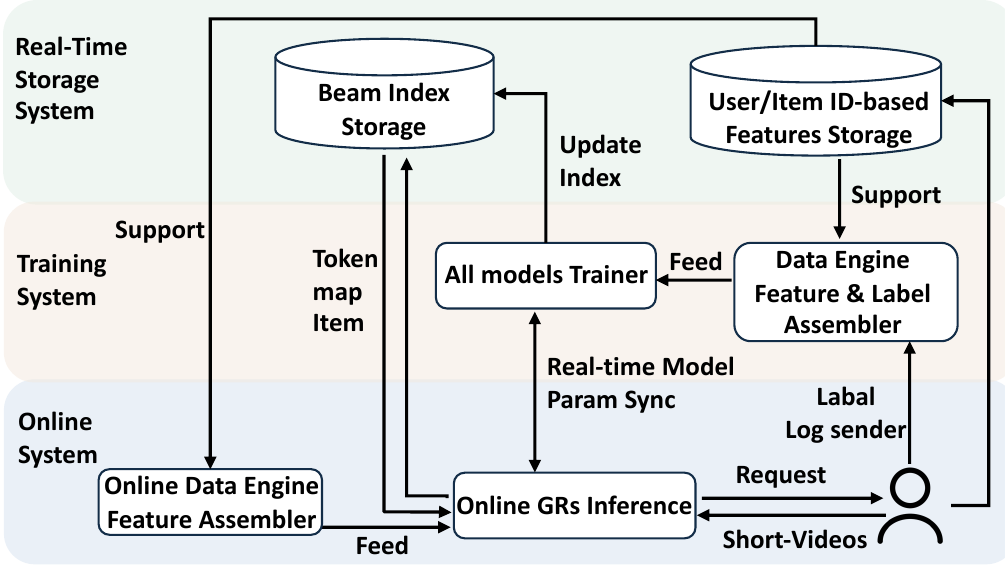}
    \caption{The overall system deployment pipeline, illustrating the interplay between models training, real-time index synchronization, and online generative recommender inference.}
    \label{fig:online deployment}
\end{figure}

As illustrated in Figure~\ref{fig:online deployment}, the PIT is deployed on Kuaishou's distributed infrastructure, comprising real-time storage, a training system, and online serving systems. During the training phase, the training engine jointly optimizes the CSA, Item-to-Token, and User-to-Token model. The latest index assignments, derived via beam search, are concurrently synchronized to the beam index, while model parameters are synchronized to the inference engine in real time. During online serving, the Data Engine aggregates user-specific features to feed the inference module. Only the generative recommender model is involved in the online prediction process.
\section{Expriments}
\label{sec:expriments}

To validate the effectiveness of our PIT framework, we conduct extensive experiments on three real-world datasets and a large-scale online production environment. Our experiments aim to answer the following research questions:
\begin{itemize}[label=$\bullet$, leftmargin=*]
    \item \textbf{RQ1:} How does PIT perform compared to representative traditional sequential and generative recommendation methods?
    \item \textbf{RQ2:} What is the significance of the key components (e.g., CSA and minimum-loss selection) to our framework?
    \item \textbf{RQ3:} How effectively does the PIT optimize codebook utilization and information entropy across different quantization layers?
    \item \textbf{RQ4:} How does PIT perform in real-world industrial scenarios compared to state-of-the-art production baselines?
\end{itemize}

\subsection{Offline Public Dataset Experiments}

\subsubsection{Datasets}
We evaluate our method on three widely used datasets from the Amazon Review 2014~\cite{he2016UpsDownsModeling} collection:``Beauty'', ``Sports and Outdoors'', and ``Toys and Games''. Following established preprocessing protocols~\cite{kang2018SelfAttentiveSequentialRecommendation,rajput2023RecommenderSystemsGenerative}, we treat all ratings as positive implicit feedback and adopt the 5-core setting, filtering out users and items with fewer than five interactions. User behavior sequences are constructed chronologically, with the maximum sequence length fixed at 50. For evaluation, we employ the widely used leave-one-out strategy~\cite{kang2018SelfAttentiveSequentialRecommendation}, utilizing the last interaction for testing, the second-to-last for validation, and the remainder for training. The detailed statistics of the processed datasets are summarized in Table \ref{tab:datasets}.

\begin{table}[h]
    \centering
    \caption{Statistics of the preprocessed datasets.}
    \label{tab:datasets}
    \resizebox{\columnwidth}{!}{
    \begin{tabular}{l|ccccc}
    \toprule
    Dataset & \#Users & \#Items & \#Interactions & Sparsity & Avg. Length \\
    \midrule
    \textbf{Beauty} & 22,363 & 12,101 & 198,502 & 99.93\% & 8.9 \\
    \textbf{Sports} & 35,598 & 18,357 & 296,337 & 99.95\% & 8.3\\
    \textbf{Toys}   & 19,412 & 11,924 & 167,597 & 99.93\% & 8.6\\
    \bottomrule
    \end{tabular}
    }
\end{table}

\begin{table*}[t]
\centering
\caption{Overall performance comparisons between the baselines and PIT on three Amazon datasets. The best results are in bold, and the second-best are underlined. Metrics: Recall (R) and NDCG (N) at $K \in \{5, 10\}$.}
\label{tab:performance}
\small
\begin{tabular}{llcccccccccccc}
\toprule
Dataset & Metric & Caser & HGN & GRU4Rec & BERT4Rec & SASRec & ReaRec & HSTU & TIGER & LETTER & ETEGRec & LC-Rec & PIT \\
\midrule
& R@5 & 0.0262 & 0.0337 & 0.0401 & 0.0226 & 0.0403 & \underline{0.0439} & 0.0423 & 0.0403 & 0.0411 & 0.0408 & 0.0433 & \textbf{0.0479} \\
Beauty & N@5 & 0.0162 & 0.0207 & 0.0260 & 0.0140 & 0.0266 & 0.0247 & 0.0282 & 0.0256 & 0.0263 & 0.0270 & \underline{0.0289} & \textbf{0.0305} \\
& R@10 & 0.0428 & 0.0554 & 0.0614 & 0.0385 & 0.0592 & \underline{0.0741} & 0.0623 & 0.0648 & 0.0654 & 0.0657 & 0.0653 & \textbf{0.0783} \\
& N@10 & 0.0216 & 0.0277 & 0.0329 & 0.0191 & 0.0327 & 0.0344 & 0.0347 & 0.0336 & 0.0340 & 0.0344 & \underline{0.0360} & \textbf{0.0403} \\
\midrule
& R@5 & 0.0127 & 0.0192 & 0.0194 & 0.0102 & 0.0203 & 0.0190 & 0.0236 & 0.0262 & 0.0266 & 0.0269 & \underline{0.0271} & \textbf{0.0284}  \\
Sports & N@5 & 0.0083 & 0.0117 & 0.0124 & 0.0063 & 0.0133 & 0.0117 & 0.0157 & 0.0165 & 0.0167 & 0.0164 & \underline{0.0177} & \textbf{0.0186} \\
& R@10 & 0.0211 & 0.0325 & 0.0320 & 0.0168 & 0.0320 & 0.0321 & 0.0348 & 0.0418 & \underline{0.0439} & 0.0435 & 0.0428 & \textbf{0.0462} \\
& N@10 & 0.0110 & 0.0160 & 0.0164 & 0.0084 & 0.0171 & 0.0158 & 0.0193 & 0.0215 & 0.0223 & 0.0221 & \underline{0.0228} & \textbf{0.0243} \\
\midrule
& R@5 & 0.0141 & 0.0356 & 0.0358 & 0.0194 & 0.0424 & \underline{0.0478} & 0.0405 & 0.0352 & 0.0383 & 0.0391 & 0.0466 & \textbf{0.0516} \\
Toys & N@5 & 0.0084 & 0.0216 & 0.0234 & 0.0120 & 0.0282 & 0.0262 & 0.0283 & 0.0220 & 0.0239 & 0.0245 & \underline{0.0323} & \textbf{0.0328}\\
& R@10 & 0.0233 & 0.0565 & 0.0571 & 0.0302 & 0.0609 & \underline{0.0771} & 0.0582 & 0.0562 & 0.0599 & 0.0602 & 0.0687 & \textbf{0.0839} \\
& N@10 & 0.0114 & 0.0284 & 0.0302 & 0.0155 & 0.0342 & 0.0357 & 0.0339 & 0.0287 & 0.0308 & 0.0321 & \underline{0.0394} & \textbf{0.0432} \\
\bottomrule
\end{tabular}
\end{table*}

\subsubsection{Baselines}
We compare PIT with a wide range of competitive baselines, categorized into traditional sequential recommendation models and recent generative recommendation models:

\textbf{Traditional Sequential Models:}
\begin{itemize}[label=$\bullet$, leftmargin=*]
\item \textbf{Caser~\cite{tang2018PersonalizedTopNSequential}}: Proposes a convolution-based approach that views the embedding matrix as an image to capture high-order Markov chains via horizontal and vertical filters.
\item \textbf{HGN~\cite{ma2019HierarchicalGatingNetworks}}: Introduces a hierarchical gating architecture effectively capturing both long-term and short-term user interests through feature gating and instance gating layers.
\item \textbf{GRU4Rec~\cite{hidasi2016SessionbasedRecommendationsRecurrent}}: A pioneering session-based method that utilizes Gated Recurrent Units (GRU) to model user clicking behaviors in sequential data.
\item \textbf{BERT4Rec~\cite{sun2019BERT4RecSequentialRecommendation}}: Adapts the deep bidirectional Transformer architecture to recommendation, utilizing Cloze tasks (Masked Language Modeling) to capture bidirectional context.
\item \textbf{SASRec~\cite{kang2018SelfAttentiveSequentialRecommendation}}: Leverages a unidirectional self-attention mechanism to identify relevant historical interactions and capture long-term sequential dependencies efficiently.
\item \textbf{ReaRec~\cite{tang2025ThinkRecommendUnleashing}}: Enhances sequential models by performing implicit reasoning on latent user intents, moving beyond simple pattern matching.
\end{itemize}
\noindent\textbf{Generative Recommendation Models:}
\begin{itemize}[label=$\bullet$, leftmargin=*]
\item \textbf{HSTU~\cite{zhai2024ActionsSpeakLouder}}: A high-performance sequential transducer designed for industrial scale, optimizing attention mechanisms for faster training and inference without compromising accuracy.
\item \textbf{TIGER~\cite{rajput2023RecommenderSystemsGenerative}}: Formulates recommendation as a generative retrieval task, utilizing a hierarchical RQ-VAE to generate semantic identifiers for items.
\item \textbf{LETTER~\cite{wang2025LearnableItemTokenization}}: Focuses on optimizing item tokenization by incorporating semantic, collaborative, and diversity regularization into a learnable framework.
\item \textbf{ETEGRec~\cite{liu2025GenerativeRecommenderEndtoEnd}}: Proposes an end-to-end training strategy with alternating optimization, introducing auxiliary recommendation-oriented losses to align representations effectively.
\item \textbf{LC-Rec~\cite{zheng2024AdaptingLargeLanguage}}: Investigates the adaptation of Large Language Models for recommendation, integrating collaborative semantics into the generation process via vector quantization.
\end{itemize}

\subsubsection{Evaluation Metrics}
We adopt two standard metrics for evaluation: \textbf{Recall@$K$} and \textbf{NDCG@$K$}. To provide a fine-grained analysis of model performance at different ranking positions, we report results for $K \in \{5, 10\}$ following~\cite{rajput2023RecommenderSystemsGenerative}. 

\subsubsection{Implementation Details}

For a fair comparison, we uniformly employ a frozen Llama3-8B~\cite{grattafiori2024Llama3Herd} to extract item embeddings for codebook construction in TIGER~\cite{rajput2023RecommenderSystemsGenerative}, LETTER~\cite{wang2025LearnableItemTokenization}, LC-Rec~\cite{zheng2024AdaptingLargeLanguage} and PIT. Notably, for TIGER and LETTER, we adopt the implementation provided by LETTER; For LC-Rec, while the embeddings follow the standardized extraction process, the subsequent fine-tuning stage is conducted using LLaMA-7B~\cite{touvron2023LLaMAOpenEfficient}, consistent with its original architecture. Furthermore, they share an identical quantization configuration using RQ-VAE~\cite{lee2022AutoregressiveImageGeneration, rajput2023RecommenderSystemsGenerative} with 4 codebooks, each containing 256 codes of dimension 32. Traditional recommendation baselines are implemented by an open-source recommendation framework RecBole~\cite{zhao2021RecBoleUnifiedComprehensive,zhao2022RecBole20More}. For ReaRec~\cite{tang2025ThinkRecommendUnleashing}, HSTU~\cite{zhai2024ActionsSpeakLouder} and ETEGRec~\cite{liu2025GenerativeRecommenderEndtoEnd}, we utilize their official implementations. Our framework employs a T5-based~\cite{raffel2023ExploringLimitsTransfer} User-to-Token model with 4 layers, $d_{model}=128$, $d_{ff}=1024$, 6 heads and a decoder-only Transformer Item-to-Token model with 4 layers, $d_{model}=64$, $d_{ff}=256$, 4 heads. We train the models for up to 200 epochs using AdamW~\cite{loshchilov2019DecoupledWeightDecay} with a weight decay of 0.01, applying early stopping based on validation NDCG@10. Unlike production environments that leverage rich real-time data for continuous Item-to-Token updates, our offline setting utilizes limited, small-scale data. To prevent instability caused by frequent updates under data scarcity, we freeze the Item-to-Token during the subsequent dynamic epoch phase, focusing exclusively on User-to-Token updates to ensure model stability. The learning rates for the User-to-Token and Item-to-Token models were tuned within the ranges of $\{5\text{e-}4, 3\text{e-}4, 1\text{e-}4\}$ and $\{1\text{e-}3, 5\text{e-}4, 1\text{e-}4\}$, respectively. We train all experiments on 8 NVIDIA RTX A800 GPUs.

\subsection{Overall Offline Performance (RQ1)}

The results of PIT and different baselines on three datasets are shown in Table ~\ref{tab:performance}. Based on the results, we have the following observations:
\begin{itemize}[label=$\bullet$, leftmargin=*]
\item PIT consistently achieves the best performance across all datasets and evaluation metrics. This significant improvement validates the effectiveness of our proposed framework in capturing complex user-item patterns even in offline scenarios with limited data. The superior performance can be attributed to the incorporation of collaborative signals, the multi-faceted representation of beam index, and the architectural superiority of the proposed model.
\item Traditional sequential recommendation methods (e.g., Caser, HGN, and GRU4Rec) generally underperform compared to most generative approaches. This performance gap is likely due to their reliance on discrete, one-hot identifier representations, which fail to capture the underlying semantic relationships between items. While models like SASRec show competitive performance among traditional baselines by leveraging self-attention mechanisms, they still struggle to match the performance of advanced generative models that utilize hierarchical semantic structures.
\item Generative recommendation models (e.g., TIGER and LC-Rec) typically show better results than traditional sequential models. Specifically, models that incorporate RQ-VAE for semantic tokenization like TIGER and LETTER demonstrate the advantage of leveraging fine-grained, hierarchical item representations. Notably, the superior performance of LETTER and ETEGRec over the standard TIGER framework underscores the critical necessity and benefits of integrating collaborative signals and end-to-end optimization. PIT further extends this advantage by better balancing the tokenization process and the recommendation task, allowing for a more robust mapping between user preferences and the generative target space even under the constraints of offline data.
\end{itemize}

\vspace{-5pt}

\begin{table}[!t]
    \centering
    \caption{Ablation study of PIT on Beauty as well as Toys and Games datasets.}
    \label{tab:ablation}
    \small
    \resizebox{0.75\columnwidth}{!}{
    \begin{tabular}{l|cc|cc}
    \toprule
    \multirow{2}{*}{\textbf{Variant}} & \multicolumn{2}{c|}{\textbf{Beauty}} & \multicolumn{2}{c}{\textbf{Toys}} \\
    & R@10 & N@10 & R@10 & N@10 \\
    \midrule
    \textbf{PIT (Full)} & \textbf{0.0783} & \textbf{0.0403} & \textbf{0.0839} & \textbf{0.0432} \\
    w/o CSA & 0.0738 & 0.0379 & 0.0771 & 0.0397 \\
    w/o MLS & 0.0742 & 0.0387 & 0.0798 & 0.0405 \\
    w/o BI & 0.0666 & 0.0349 & 0.0622 & 0.0321 \\
    w/o MLS \& BI & 0.0606 & 0.0317 & 0.0498 & 0.0252 \\
    TIGER  & 0.0648 & 0.0336 & 0.0562 & 0.0287 \\
    \bottomrule
    \end{tabular}
    }
    \vspace{-15pt}
\end{table}

\begin{table*}[ht]
\centering
\caption{Online A/B test results: Relative improvement of OneRec-V2-PIT over OneRec-V2 on Kuaishou and Kuaishou Lite.}
\label{tab:online}
\resizebox{\textwidth}{!}{
\begin{tabular}{l|cccc|ccccc}
\toprule
\multirow{2}{*}{\textbf{Scenarios}} & \multicolumn{4}{c|}{\textbf{Important Online Metrics}} & \multicolumn{5}{c}{\textbf{Interaction Metrics}} \\
\cmidrule(lr){2-5} \cmidrule(lr){6-10}
 & \textbf{APP Stay Time} & \textbf{LT7} & \textbf{Watch Time} & \textbf{Video View} & \textbf{Like} & \textbf{Follow} & \textbf{Comment} & \textbf{Collect} & \textbf{Forward} \\
\midrule
\textbf{Kuaishou}      & \textbf{+0.217\%} & \textbf{+0.056\%} & \textbf{+0.371\%} & \textbf{+0.239\%} & +0.652\% & +0.674\% & +0.130\% & +0.131\% & +0.073\% \\
\textbf{Kuaishou Lite} & \textbf{+0.402\%} & \textbf{+0.146\%} & \textbf{+0.531\%} & \textbf{+1.664\%} & +0.728\% & +0.070\% & +0.811\% & +0.042\% & +0.199\% \\
\bottomrule
\end{tabular}
}
\end{table*}

\subsection{Ablation Studies (RQ2)}
To investigate the individual contribution of each component within PIT, we conduct a comprehensive ablation study on the Amazon Beauty as well as Toys and Games. The results, summarized in Table \ref{tab:ablation}, compare the full PIT model against four variants
\begin{itemize}[label=$\bullet$, leftmargin=*]
    \item \textbf{w/o CSA}: Removes the collaborative signal alignment (CSA), thereby decoupling the global collaborative semantics from the item-to-token generative process. The resulting consistent performance decay validates that aligning collaborative signals through shared item embeddings is essential for bridging the semantic gap.
    
    \item \textbf{w/o MLS}: Replaces the minimum-loss selection (MLS) mechanism with a stochastic selection strategy, neutralizing the optimal alignment between beam candidates and user preferences. The evident superiority of PIT(Full) over this variant indicates that loss-based selection effectively filters out generative noise.
    
    \item \textbf{w/o BI}: Constrains the model to a rigid one-to-one mapping by disabling the beam index (BI) functionality, effectively reverting to a unique SID assignment. The sharp performance decline observed here highlights that multi-view indexing is fundamental for capturing complex item semantics.
    
    \item \textbf{w/o MLS \& BI}: A minimalist configuration where Item-to-Token reduces to a generative tokenizer equipped only with standard RQ-VAE capabilities. The substantial gap between PIT(Full) and this variant indicates that without these modules, the process degrades into a suboptimal tokenizer. Furthermore, the fact that this variant even underperforms the original TIGER underscores the critical role of MLS and BI; without them, the dynamic generative indexing fails to match the efficacy of traditional static tokenizer.
\end{itemize}

\begin{figure}[h]
    \centering
    \begin{subfigure}{\linewidth}
        \centering
        \includegraphics[width=\linewidth]{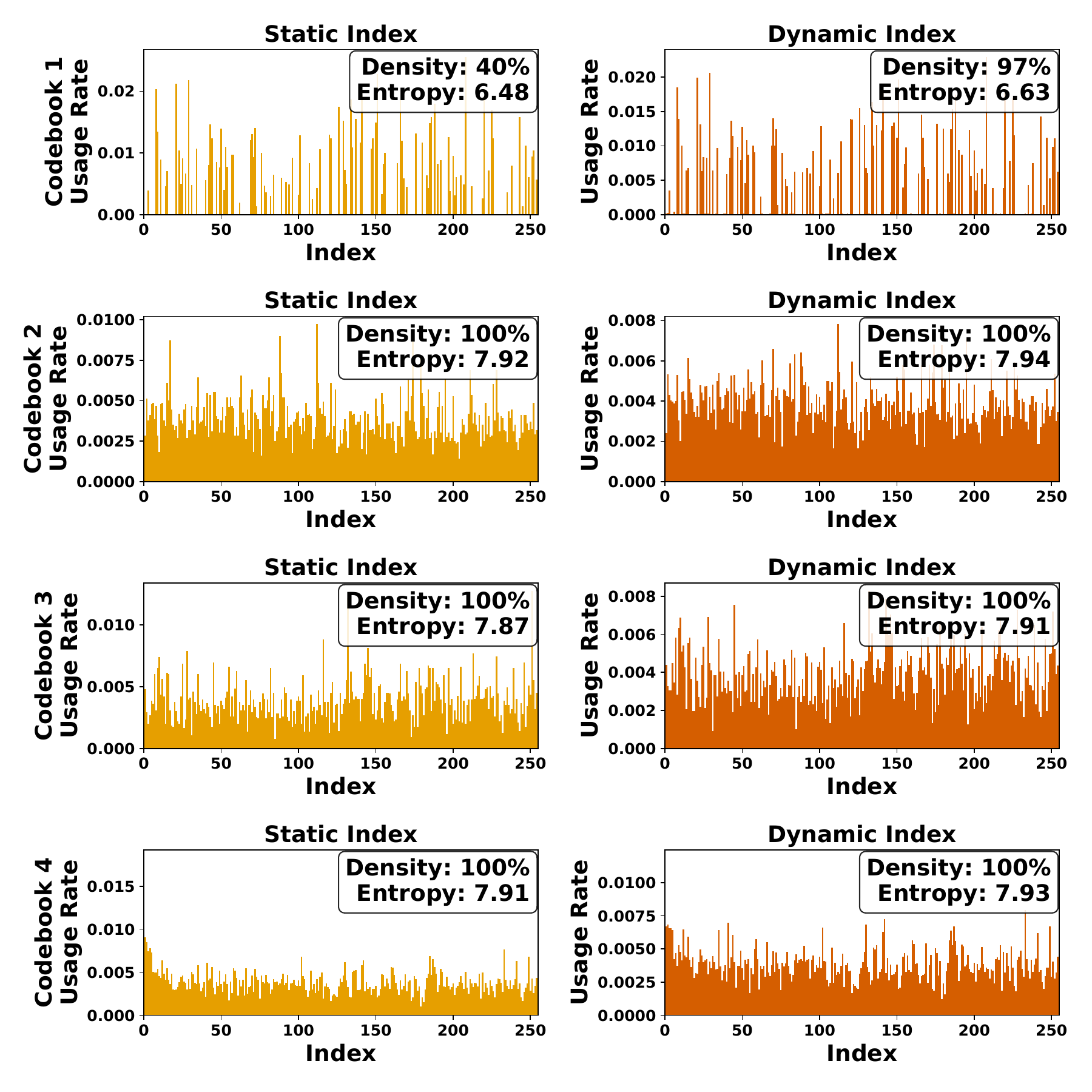}
        \caption{Codebook Entropy Analysis on Toys and Games}
        \label{fig:compare_codebook_entropy}
    \end{subfigure}
    \begin{subfigure}{\linewidth}
        \centering
        \includegraphics[width=\linewidth]{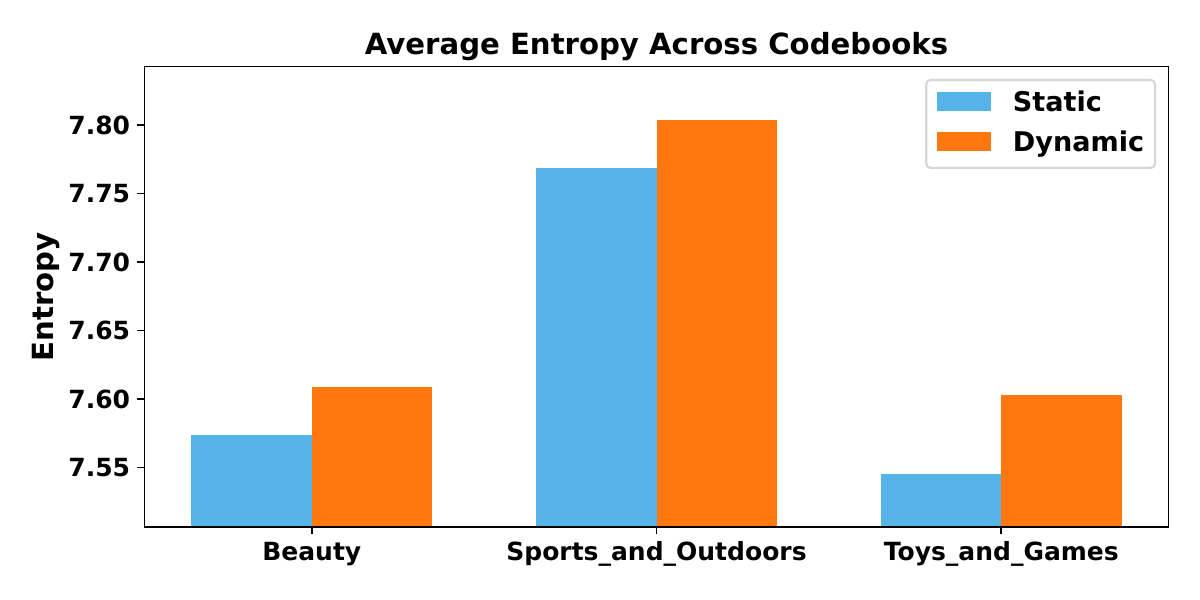}
        \caption{Total Entropy Comparison on three datasets}
        \label{fig:compare_total_entropy}
    \end{subfigure}
    \caption{Quantitative analysis of codebook utilization. (a) Codeword distribution across quantization layers on the Toys and Games dataset. (b) Overall comparison of average entropy across three datasets.}
    \label{fig:compare_codebook_entropy}
\end{figure}

\subsection{Codebook Entropy Analysis (RQ3)}
To evaluate the discrete codebook’s efficiency, we analyzed its density as a measure of the ratio of active to total codewords and Shannon entropy to assess the uniformity of codeword distribution across quantization layers using the Toys and Games dataset. Additionally, we compared the average entropy across all codebooks for the three datasets, as illustrated in Figure~\ref{fig:compare_codebook_entropy}. The empirical results demonstrate that PIT components significantly enhance the utilization of the initial quantization layers. While the overall improvement in entropy appears moderate, we attribute this constraint primarily to the fact that the Item-to-Token model was not fully updated in a synchronous manner due to the scale limitations of the offline datasets. Despite these challenges, our experiments demonstrate that the proposed components effectively optimize entropy distribution and increase codebook density.

\subsection{Online A/B Result (RQ4)}

To verify the effectiveness of PIT in a real-world industrial environment, we deployed it on a large-scale short video recommendation platform. We conducted experiments over a 24-day period, spanning two major recommendation scenarios at Kuaishou. We compare PIT against the current production baseline OneRec-V2~\cite{luo2024QARMQuantitativeAlignment, zhou2025OneRecV2TechnicalReport}). The results are reported in Table ~\ref{tab:online}. The proposed model demonstrates superior performance across all metrics in both scenarios, validating the effectiveness of our approach. Specifically, regarding the online metrics in our internal scenarios, we prioritize Stay Time and LT7 (Lifetime over 7 days). These metrics function as the primary indicators for assessing user stickiness and long-term retention, respectively. Furthermore, the beam index is updated synchronously during streaming training, allowing our personal tokenizer to be seamlessly integrated into various generative recommender architectures. These models, irrespective of their scale or structure, can directly leverage the dynamic index by retrieving item-to-token mappings (details in Appendix~\ref{app:online_res_explain}).

\section{Conclusion}
\label{sec:conclusion}

In this study, we presented a dynamic end-to-end generative retrieval framework specifically designed to achieve a deeper integration of collaborative signals and shortcomings of tokenizer in industrial scenarios. By integrating a dynamic tokenizer and beam index, our approach moves beyond the constraints of traditional fixed-token indexing. This architecture allows for a more flexible and granular representation of semantic identifiers, ensuring that the generative process is more aligned with the underlying data distribution. Empirical results on Kuaishou's Short Videos scenarios demonstrate the effectiveness of our method on large-industry Generative Recommendation. Future work involves substituting the basic DNN backbone with high-quality ranking-stage embeddings and exploring advanced integration strategies of collaborative filtering with semantic identifiers for enhanced synergistic representations.


\bibliographystyle{ACM-Reference-Format}
\bibliography{sample-base}

\appendix
\section{Appendix}
\label{sec:appendix}

\subsection{Beam Index Online}\label{app:beam_index}

A defining characteristic of Beam Index is that the index is not a static lookup table but a fluid, multi-view evolutionary topology. During the training process, the index is automatically constructed and refreshed in real-time based on the output of the Item-to-Token model.

\paragraph{\textbf{Online Storage of Beam Index}}
To accommodate Kuaishou's large-scale production environment, we use a high-performance GNN~\cite{4700287} storage service designed to facilitate bi-directional mappings between SIDs and item identifiers, as well as the retrieval of their heterogeneous side information. Specifically, each item identifier is projected onto eight distinct token nodes. Within this architecture, the ingestion timestamps and significance weights constitute the critical dimensions of the side info; these attributes dictate the definitive associations between items and tokens, thereby serving as the foundational logic for constructing the Beam Index.

\begin{algorithm}[ht]
\caption{Dynamic Beam Index Evolution and Pruning}
\label{alg:beam_index_update}
\begin{algorithmic}[1]
\STATE \textbf{Input:} Item $i$, Historical tokens $\mathcal{T}_{\text{fetch}}$, Predicted tokens $\mathcal{T}_{\text{pred}}$, Momentum $\gamma$, Index capacity $B$, Time threshold $\Delta t$
\STATE \textbf{Output:} Updated SID set $\mathcal{S}_i$

\STATE \textbf{// Step 1: Relevance Weight Calculation}
\FOR{each $\tau \in \mathcal{T}_{\text{pred}}$}
    \STATE Calculate current weight $W_{i,\tau}^{\text{cur}}$ using Eq. (\ref{equation_weight})
\ENDFOR

\STATE \textbf{// Step 2: Momentum-based Weight Update}
\STATE $\mathcal{T}_{\text{merge}} \leftarrow \mathcal{T}_{\text{fetch}} \cup \mathcal{T}_{\text{pred}}$
\FOR{each $\tau \in \mathcal{T}_{\text{merge}}$}
    \IF{$\tau \in \mathcal{T}_{\text{fetch}} \cap \mathcal{T}_{\text{pred}}$}
        \STATE $W_{i,\tau} \leftarrow \gamma \cdot W_{i,\tau}^{\text{old}} + (1-\gamma) \cdot W_{i,\tau}^{\text{cur}}$
    \ELSIF{$\tau \in \mathcal{T}_{\text{pred}} \setminus \mathcal{T}_{\text{fetch}}$}
        \STATE $W_{i,\tau} \leftarrow W_{i,\tau}^{\text{cur}}$
    \ELSE
        \STATE $W_{i,\tau} \leftarrow W_{i,\tau}^{\text{old}}$
    \ENDIF
\ENDFOR

\STATE \textbf{// Step 3: Graph Integrity Pruning}
\FOR{each $\tau \in \mathcal{T}_{\text{merge}}$}
    \IF{$\tau \in \mathcal{T}_{\text{fetch}}$}
        \STATE \COMMENT{\textbf{Forward Consistency Maintenance}}
        \IF{$\text{Map}(\tau) \neq i$}
            \STATE Terminate the mapping: $\mathcal{T}_{\text{merge}} \leftarrow \mathcal{T}_{\text{merge}} \setminus \{\tau\}$
        \ENDIF
    \ELSIF{$\tau \in \mathcal{T}_{\text{pred}}$ \AND $\text{Map}(\tau) = j$ ($j \neq i$)}
        \STATE \COMMENT{\textbf{Temporal Ownership Resolution}}
        \IF{$\text{Timestamp}(\tau \to j) < \text{Now}() - \Delta t$}
            \STATE Prune stale link $(j, \tau)$ and reassign to $i$
        \ELSE
            \STATE Respect incumbent ownership: $\mathcal{T}_{\text{merge}} \leftarrow \mathcal{T}_{\text{merge}} \setminus \{\tau\}$
        \ENDIF
    \ENDIF
\ENDFOR

\STATE \textbf{// Step 4: Top-B Consolidation}
\STATE $\mathcal{S}_i \leftarrow \text{Top-}B \text{ tokens in } \mathcal{T}_{\text{merge}} \text{ sorted by } W_{i,\tau}$
\RETURN $\mathcal{S}_i$
\end{algorithmic}
\end{algorithm}
\paragraph{\textbf{Online Update of Beam Index}}
The foundation of the Beam Index update mechanism is a quantitative measure of the representational strength between an item $i$ and a token sequence $c$. We define the relevance weight $W_{i,c}$ based on the generative confidence:
\begin{equation}\label{equation_weight}
    W_{i,c} = \alpha - \sum_{t=1}^{|c|} \mathcal{L}_{\text{CE}}(x_t | x_{<t}, i),
\end{equation}
where $\mathcal{L}_{\text{CE}}$ denotes the cross-entropy loss at each decoding step, and $\alpha$ is a normalization constant (empirically set to 100) to ensure a positive weight distribution. A higher $W_{i,c}$ indicates that token sequence $c$ captures the semantics of item $i$ with lower uncertainty. Triggered by each Item-to-Token inference step, the Beam Index undergoes a real-time refresh as shown in Algorithm~\ref{alg:beam_index_update}. We first synthesize a unified candidate set $\mathcal{T}_{\text{merge}}$ by amalgamating historically indexed tokens ($\mathcal{T}_{\text{fetch}}$) and currently predicted tokens ($\mathcal{T}_{\text{pred}}$). To balance historical stability with instantaneous generative confidence, we implement a differential weight update strategy prior to topology pruning. Specifically, the weight $W_{i,\tau}$ for each token $\tau \in \mathcal{T}_{\text{merge}}$ is determined by its intersection status:
\begin{equation}
    W_{i,\tau} = 
    \begin{cases} 
    \gamma \cdot W_{i,\tau}^{\text{old}} + (1-\gamma) \cdot W_{i,\tau}^{\text{cur}} & \text{if } \tau \in \mathcal{T}_{\text{fetch}} \cap \mathcal{T}_{\text{pred}}, \\
    W_{i,\tau}^{\text{cur}} & \text{if } \tau \in \mathcal{T}_{\text{pred}} \setminus \mathcal{T}_{\text{fetch}}, \\
    W_{i,\tau}^{\text{old}} & \text{if } \tau \in \mathcal{T}_{\text{fetch}} \setminus \mathcal{T}_{\text{pred}},
    \end{cases}
\end{equation}
where $W_{i,\tau}^{\text{cur}}$ is calculated via Eq.~\ref{equation_weight}, $W_{i,\tau}^{\text{old}}$ denotes the stored weight, and $\gamma \in [0,1]$ is a momentum coefficient controlling the update rate. This mechanism ensures that recurring indices (overlap) are reinforced through momentum~\cite{he2019MomentumContrastUnsupervised}, while novel tokens are initialized with their current generative uncertainty. To ensure graph integrity and maximize retrieval relevance under storage constraints, the topology is refined through the following protocols:
\begin{itemize}[label=$\bullet$, leftmargin=*]
    \item \textbf{Forward Consistency Maintenance:} For any historically fetched token $\tau \in \mathcal{T}_{\text{fetch}}$, we verify the global reverse index $\text{Map}(\tau)$. If $\text{Map}(\tau)$ no longer points to the current item $i$, the forward association is severed to prevent graph inconsistency.
    \item \textbf{Temporal Ownership Resolution:} For any newly predicted token $\tau \in \mathcal{T}_{\text{pred}}$ involved in a conflict (where $\text{Map}(\tau) \neq i$), we evaluate the recency of the existing link. If $\tau$ was engaged by another item within a recent time window (indicating fresh usage), we respect the incumbent ownership and exclude $\tau$. Conversely, stale links are actively pruned and reassigned to item $i$.
    
    \item \textbf{Top-$B$ Consolidation:} To strictly enforce storage constraints, the definitive index for item $i$ is consolidated by retaining only the top-$B$ tokens with the highest weights from the surviving candidates. In cases where the valid candidate count falls below $B$, all remaining tokens are preserved with padding.
\end{itemize}
Consequently, at any given moment, the set $\mathcal{S}_i$ represents the optimal, competitively selected aliases for item $i$. Upon the completion of the training phase, the co-evolutionary process terminates, and the item SIDs effectively stabilize. Our framework directly utilizes the final optimized Beam Index obtained from the last training epoch.
\subsection{Online Implementation}\label{app:implement_kuaishou}
\subsubsection{\textbf{Online Traning}}\label{app:online_training}
\begin{figure*}[ht]
    \centering
    \includegraphics[width=\textwidth]{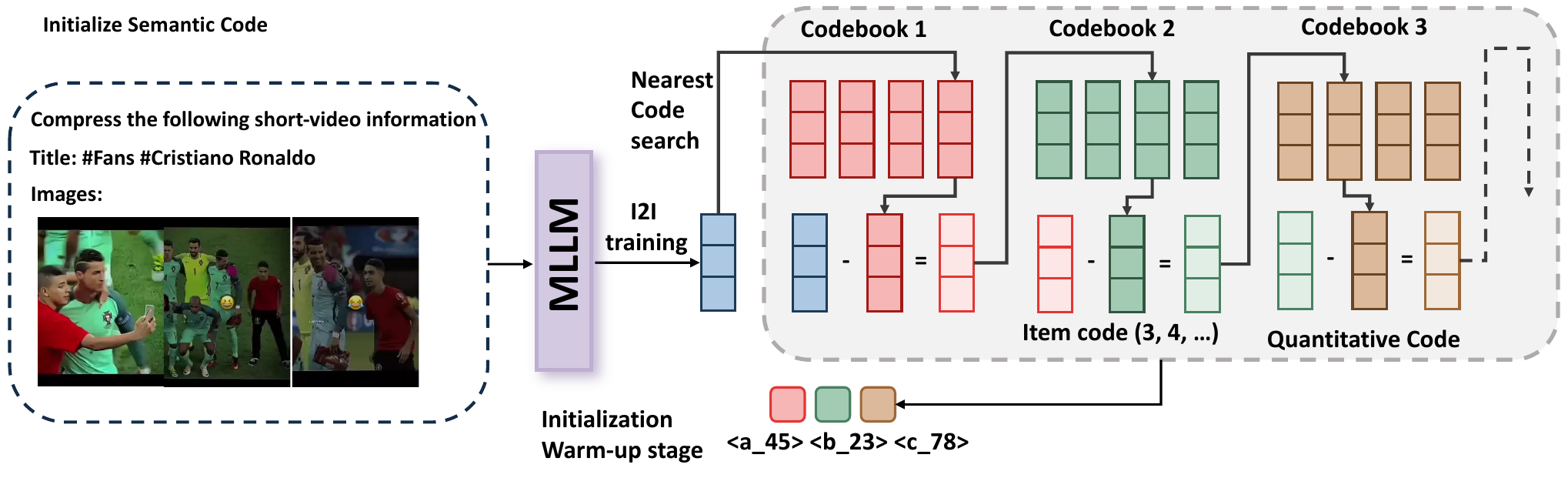}
    \caption{Architectural overview of the RQ-Kmeans initialization. The codebook is pre-trained via Item-to-Item (I2I) contrastive learning to provide a semantically rich starting point for subsequent dynamic Beam Index updates.}
    \label{fig:rqkmeans}
\end{figure*}

\paragraph{\textbf{Online Training Settings}}
For the online experiments, we strictly adhere to the configuration proposed in our methodology. The User-to-Token component is fully aligned with the deployed OneRec-V2~\cite{zhou2025OneRecV2TechnicalReport} model. For the Item-to-Token component, we employ a 0.1B parameter decoder-only Transformer. Both components share a vocabulary size of $8192 \times 3$. For the DIN~\cite{zhou2018DeepInterestNetwork} model, we utilize a simplified architecture consisting of a 4-layer MLP and a single cross-attention Transformer layer. Regarding the online Beam Index in Appendix~\ref{app:beam_index}, we configure the index capacity to 8 for each item, such that each item identifier is represented by 8 distinct tokens. It is noteworthy that during the warmup phase, the Beam Index is continuously populated leveraging a codebook initialized via RQ-Kmeans, which is pre-trained through Item-to-Item (I2I) contrastive learning~\cite{luo2024QARMQuantitativeAlignment}. The detailed architectural initialization and the generation of semantic tokens via RQ-Kmeans are illustrated in Figure~\ref{fig:rqkmeans}. 

\paragraph{\textbf{Details of Online Training}}
(1) \textit{Warm-up:} The curriculum learning~\cite{bengio2009CurriculumLearning} begins with a warm-up objective, where the Beam Index is constructed by inserting deterministic, one-to-one mappings between SIDs and their corresponding item identifiers. This phase serves as a structural initialization, ensuring the model captures the fundamental item-token alignment. Throughout this period, the User-to-Token loss trajectory remains consistent with the OneRec-V2 baseline. The transition to the subsequent phase is triggered once the model's performance aligns with the convergence level of OneRec-V2, indicating that the model has sufficiently mastered the basic latent representations. (2) \textit{Dynamic Evolution:} Upon switching, the model enters a more challenging dynamic learning objective. While a transient loss spike occurs due to the distributional shift of the index, the loss eventually stabilizes slightly above the static baseline. In this phase, the Beam Index evolves from rigid assignments to the momentum-weighted, one-to-many associations as detailed in Appendix~\ref{app:beam_index}. 
Specifically, for the User-to-Token objective, we implement a minimum-loss selection mechanism to refine the labels. For the Item-to-Token objective, we introduce a gradient-based filtering strategy: updates are selectively executed only when the associated User-to-Token label yields a weight exceeding 90. Within our $8192 \times 3$ vocabulary space, we posit that an accumulated loss within a threshold of 10 signifies a statistically valid signal rather than a noisy sample, thereby ensuring the precision and stability of the dynamic index evolution.

\subsubsection{\textbf{Online Experiments And Index Reuse}}\label{app:online_res_explain}
\begin{figure}[h]
    \centering
    \includegraphics[width=\linewidth]{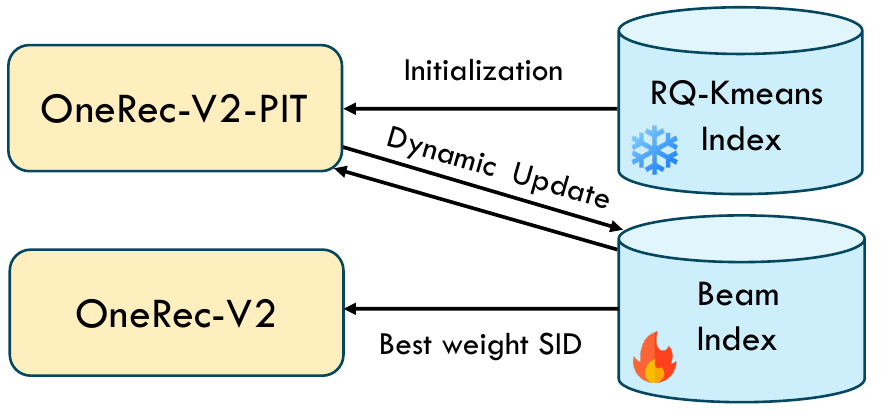}
    \caption{PIT Index Transfer Framework: Enabling Universal Index Reuse Across Generative Recommenders}
    \label{fig:index_reuse}
\end{figure}

To evaluate the practical utility of PIT in real-world industrial systems, we conducted extensive online A/B tests on two large-scale short-video platforms: Kuaishou and Kuaishou Lite. For each application, 5\% of the total traffic was randomly allocated to the experiment group (equipped with PIT-enhanced OneRec-V2) and the control group (base OneRec-V2), respectively. Notably, while PIT model was trained on Kuaishou Lite data, its Beam Index was directly transferred to the OneRec-V2 training on Kuaishou as shown in A/B tests. Specifically, the system selects the index with the highest learned weight to guide the feature representation in OneRec-V2, which is shown in figure~\ref{fig:index_reuse}. The observed performance gap between the two platforms suggests a potential distribution shift, as the index was fully optimized for Kuaishou Lite. To ensure long-term robustness and account for the stability issues discussed in Appendix~\ref{app:limitation_resolve}, the experiments were monitored over a 24-day duration. Results demonstrate that PIT significantly outperforms the baseline across core engagement metrics, including App Stay Time, LT7 (Lifetime over 7 days), Watch Time, and Video Views. However, a relatively moderate improvement was observed in interaction-based metrics (e.g., Collect, Forward) compared to the primary retention indicators. Notably, the persistent improvement in LT7 demonstrates the long-term user retention capability of our proposed method, which is often considered a challenging objective in industrial recommendation systems.
It is worth highlighting that our online inference architecture remains highly efficient. Although multiple models and iterative augmentation are employed during the offline training phase, only the User-to-Token model is required for real-time serving. Since the tokenizer and collaborative signals alignment components are strictly confined to the training stage, the inference path is decoupled from these complex processes. Consequently, our approach maintains a service latency consistent with the standard OneRec-V2, ensuring seamless integration into production environments without additional computational overhead.

\subsection{Empirical Findings and Solutions}\label{app:limitation_resolve}
During continuous multi-week streaming updates, we observed a significant index decay phenomenon. As training extends over several weeks, the item embeddings $X_i$ within the CSA module increasingly converge toward collaborative signals, which inadvertently leads to them being dominated by popularity biases. Consequently, the Item-to-Token model ($\phi$) overfits to this biased distribution, leading to a phenomenon where distinct items are mapped to a narrow set of ``hot'' token sequences, thereby reducing representational diversity and retrieval precision. To counteract this instability and maintain the semantic integrity of the learned identifiers, we introduce a \textit{Reference-Regularized Constraint} mechanism, drawing inspiration from the Kullback-Leibler (KL) divergence regularization used in Group Relative Policy Optimization~\cite{shao2024DeepSeekMathPushingLimits}. 

\paragraph{\textbf{Reference Model Construction.}} We introduce a reference tokenizer, parameterized by $\phi_{ref}$, which shares the same architecture as the Item-to-Token model. Unlike the primary tokenizer $\phi$ which evolves with user feedback, $\phi_{ref}$ is trained to maintain a stationary mapping between item attributes and RQ-Kmeans SIDs. It is supervised by the initial collaborative-semantic identifiers $\bar{c}_i$ established during the all phase (Section~\ref{sec:learning}):
\begin{equation}
    \mathcal{L}_{ref} = -\sum_{l=1}^{L} \log P_{\phi_{ref}}(\bar{c}_{i,l} | \bar{c}_{i,<l}, \text{sg}(X_i)).
\end{equation}
Crucially, $\phi_{ref}$ serves as a stable anchor. It captures the fundamental semantic-collaborative mapping without being corrupted by the dynamic shifts in popularity-driven user feedback during the co-evolution phase.

\paragraph{\textbf{KL-Divergence Constraint.}} To ensure that the dynamic Item-to-Token model $\phi$ does not deviate excessively from the stable semantic space, we impose a KL-divergence constraint between the output distributions of the current model and the reference model. For each generation step $l$, the constraint is defined as:
\begin{equation}
    \mathbb{D}_{KL}(P_{\phi} \| P_{\phi_{ref}}) = \frac{1}{L} \sum_{l=1}^{L} \left( \frac{P_{\phi_{ref}}}{P_{\phi}} - \log \frac{P_{\phi_{ref}}}{P_{\phi}} - 1 \right),
\end{equation}
where $P_{\phi}$ and $P_{\phi_{ref}}$ denote the shortcut for $P_{\phi}(c^*_{i,l} | c^*_{i,<l}, \text{sg}(X_i))$ and $P_{\phi_{ref}}(c^*_{i,l} | c^*_{i,<l}, \text{sg}(X_i))$, respectively.

\paragraph{\textbf{Regularized Joint Objective.}} Incorporating this constraint, the total optimization objective for the framework in the co-evolution phase is reformulated to balance user-guided evolution and semantic stability:
\begin{equation}
    \mathcal{L}_{total} = \mathcal{L}_{dynamic} + \alpha \left( \mathcal{L}_{ref}(\bar{c}_{i}) + \eta \mathbb{D}_{KL}(P_{\phi} \| P_{\phi_{ref}}) \right),
\end{equation}
where $\eta$ is a scaling hyperparameter. This regularization prevents the tokenizer from being overly biased by popularity signals, which would otherwise degrade the performance of highly popular items, ensuring that semantic nuances are preserved even as collaborative embeddings drift towards high-heat patterns.

\paragraph{\textbf{Future Outlook.}} While the reference-regularized mechanism provides a robust safeguard against popularity-induced index collapse, it remains a grounded approach relying on static anchors. Future research will focus on developing more adaptive, self-correcting update strategies that ensure long-term index stability without a fixed reference, alongside investigating deeper structural synergies between collaborative filtering embeddings and semantic identifiers. We invite the community to explore even more sophisticated and innovative paradigms to further push the boundaries of generative recommendation and index evolution.

\end{document}